\documentclass[12pt]{iopart}
\usepackage{iopams}
\usepackage{color}
\usepackage{hyperref}
\hypersetup{colorlinks=true,
            citecolor=blue,
            urlcolor=blue,
            colorlinks=blue}
\usepackage{graphicx}            
\newcommand{\binom}[2]{\left(\begin{array}{@{}c@{}} #1 \\ #2 \end{array}\right) }

\begin{document}

\markboth{Sergio Floquet, Marco Trindade and David Vianna}{Lie Algebras and Generalized Thermal Coherent States}

\title[Lie Algebras and Generalized Thermal Coherent States]{Lie Algebras and Generalized Thermal Coherent States}

\author{\hspace{-2.4cm} Sergio Floquet$^{1,2,\dag}$, Marco A. S. Trindade$^{3,\ddag}$ and J. David M. Vianna$^{2,4,\S}$}

\address{ \hspace{-2.2cm} $^{1}${Colegiado de Engenharia Civil, UNIVASF, Juazeiro-BA, 48.902-300, Brasil}\\  
\hspace{-2.2cm} $^{2}${Instituto de F\'isica, UFBA, 40.170-115, Salvador-BA, Brasil}\\ 
\hspace{-2.2cm} $^{3}${Departamento de Ci\^encias Exatas e da Terra - Campus I, UNEB, Salvador-BA, 41.150-000, Brasil}\\ 
\hspace{-2.2cm} $^{4}${Instituto de F\'isica, UnB, Bras\'ilia-DF, 70.919-970, Brasil} \\
\hspace{-2.2cm} $^{\dag}$ sergio.floquet@univasf.edu.br,  $^{\ddag}$ mtrindade@uneb.br,   $^{\S}$ jdavid@fis.unb.br }


\begin{abstract}
In this paper, we developed an algebraic formulation for the generalized thermal coherent states with a Thermofield Dynamics approach for multi-modes, based on coset space of Lie groups. In particular, we applied our construction on $SU(2)$ and $SU(1,1)$ symmetries and we obtain their thermal coherent states and density operator. We also calculate their thermal quantum Fidelity and thermal  Wigner function. 
\end{abstract} 

  
\section{Introduction}

The concept  of coherent states was introduced by Schr\"odinger \cite{schroedinger26}  in 1926, associated with classical states of the quantum harmonic oscillator. In 1963, Glauber\cite{glauber1,glauber2,glauber3}  coined the term coherent state and showed that it is adequate to describe a coherent laser beam in Quantum Optics. At the same time, Klauder\cite{klauder63} presents a generalization through an over-completeness property.
A group-theoretical formulation for the generalized coherent states was carried by Perelomov\cite{perelomov72,perelomov86} and Gilmore\cite{gilmore72} independently. According to this construction, if $G$ is a Lie group and $H$ is the isotropy subgroup for the state $\vert \psi_{0} \rangle  \in \mathcal{H}$ (Hilbert space), the coherent states are defined by a generalized displacement operator on $\vert \psi_{0} \rangle$, where there is one-to-one correspondence to the coset representation of
 $G / H$. A beautiful review can be encountered in the reference \cite{zhang90}.

In context of Thermofield Dynamics (TFD), that is a real-time quantum field theory at finite temperature, a thermalized version of field coherent states was introduced by Khanna \emph{et al} \cite{khanna}.  A myriad  of applications of TFD  has been developed in Quantum Optics \cite{barnett89,barnett85}, Cosmology and String Theory \cite{israel76,maldacena03}, Gauge Theory \cite{ojima81}, Casimir effect \cite{belich11},  Quantum Dissipation,\cite{celeghini1} Quantum Entanglement \cite{santana02} and Quantum Information \cite{masaki14}. 

 The TFD formulation, or other formalism based in the duplication of the degrees of freedom, are natural candidates to be described by Hopf algebra\cite{celeghini2,celeghini3}. 
 TFD can be represent by a \textit{q-deformed} Hopf algebra, allowing a classification for the unitary inequivalent representation on Quantum Field Theory. Our interesting relies on another  algebraic approach based in Lie algebras representation, emphasized observables and generators symmetry\cite{santana00,santana95}. 
This approach is interesting for the generalized coherent states since it is provides a general prescription to define thermalized states based in representation of Lie algebras and the pure states. 

The main purpose of this paper is to present a general formulation for the generalized thermal coherent states  for multi-modes, based on coset spaces of Lie groups, allowing to explore the symmetries of any coherent state of a  Lie algebras in a thermal scenario. 
In order to illustrate our formalism, we consider the symmetries of $\mathfrak{su}(2)$  and $\mathfrak{su}(1,1)$ Lie algebras in this scenario. We then calculate the thermal density operators. This calculation allows to obtain quantities of interest for quantum information such as quantum Fidelity and Wigner function. In this case these quantities are explicitly dependent on the temperature, and we denote by thermal quantum Fidelity and thermal Wigner function, respectively.
%


The structure of the paper is as follows: in Section \ref{sec2}, we first review the construction of Perelomov and Gilmore, bringing as examples the cases of the Harmonic Oscillator, the compact $\mathfrak{su}(2)$ and the non-compact $\mathfrak{su}(1,1)$ Lie algebras. Section \ref{sec3} we present the formalism of TFD used to analysis the thermal effects, based in the duplication of the Hilbert space. Section \ref{sec4} is devoted to the main propose of this paper, that is we present the construction of the Thermal Coherent State for a arbitrary Lie algebra with multi-modes. In Section \ref{sec5} we apply our construction for the Thermal Coherent State of the $\mathfrak{su}(2)$ and $\mathfrak{su}(1,1)$, obtain the Thermal Density Operator, the Thermal Fidelity and the Thermal Wigner Function.

\section{Coherent State for a Arbitrary Lie Algebra}\label{sec2}

In 1972 Perelomov\cite{perelomov72} and Gilmore\cite{gilmore72} independently show that the more consistent form to construct Coherent States for a arbitrary Lie algebra is by generalizing the concept of Displacement Operator and develope a group-theoretical approach. Let $\mathbb{H}$ be the Hamiltonian of the system, with a symmetry group $G$, that for us is a Lie group, with Lie algebra $\mathfrak{g}$ and Hilbert space given by $\mathcal{H}$, then

\begin{itemize}
  \item[$i)$] If we define the state $\vert \psi_{0} \rangle \in \mathcal{H}$ as a reference state, the maximum stability subgroup, that will be denoted for $H$, is a subgroup of $G$ that consists of all group elements that leave the reference state invariant, that is,
\begin{eqnarray}
   h \vert \psi_{0}  \rangle & = & \exp[  i\phi (h) ] \vert \psi_{0} \rangle , \ \ h \in H  .
\end{eqnarray}

\item[$ii)$] The coset space $G/H$, with every element of $g \in G$ have a unique decomposition into a product of two group elements
\begin{eqnarray}
  g & = & \Omega h,  \ \ g\in G, \ \ h\in H, \ \ \Omega \in G/H,
\end{eqnarray}
so for every reference state $\vert \psi_{0} \rangle$ we can obtain a unique coset space. An action of an arbitrary group element $g\in G$ on $\vert \psi_{0} \rangle$ is given by
\begin{eqnarray}
g \vert \psi_{0} \rangle & = &  \Omega \vert \psi_{0} \rangle  e^{i\phi(h)}.
\end{eqnarray}

\item[$iii)$] The definition of the Coherent State for a arbitrary Lie algebra is given by the combination
\begin{eqnarray}
  \vert \Lambda, \Omega \rangle & = & \Omega \vert \psi_{0} \rangle,
\end{eqnarray}
where $\Omega$ is the generalization of the Displacement Operator, which can be rewritten in terms of elements of the Lie algebra $\mathfrak{g}$.
\end{itemize}

\subsection{Coherent State for the Harmonic Oscillator}

The usual Hamiltonian of the Harmonic Oscillator is given by 
\begin{eqnarray}
  \mathbb{H} & = & \omega \left( a^{\dagger} a +\frac{1}{2} \right),
\end{eqnarray}
with $\omega$ is the frequency, $\hbar=1$, $a^{\dagger}$ and $a$ are the creation and annihilation operators respectively, satisfying 
\begin{eqnarray}
  \left[a, a^{\dagger} \right] & = & I, \ \ \left[ a, I \right] \ = \ 0, \ \ \left[ a^{\dagger}, I \right] \ = \ 0, \label{weyl01}
\end{eqnarray}
where $I$ is unit operator and $\left[,\right]$ is the usual commutation relation.
 
 Consider the Hamiltonian states as $\left\lbrace \vert n \rangle \right\rbrace_{n  \in  \mathbb{N}}$, that form a basis for a Hilbert space $\mathcal{H}$.
  The set of operators $\left\lbrace a^{\dagger} a,  a^{\dagger}, a, I \right\rbrace$ spans a Lie algebra that is, denoted by $\mathfrak{w}_{1}$. The associated Lie group is $W_{1}$, Heisenberg-Weyl group. The corresponding Hilbert space for $W_{1}$ is spanned by eigenstates $\vert n \rangle$.
 The maximum stability subgroup is the $U(1) \otimes U(1)$  so that 
  \begin{eqnarray}
  e^{\alpha a^{\dagger} - \alpha^{*} a}  & \in &    W_{1} / \left[ U(1) \otimes U(1) \right], \ \  \alpha \in \mathbb{C}  .
  \end{eqnarray}

  The Coherent State of the Harmonic Oscillator is
  \begin{eqnarray}
  \vert \alpha \rangle & = & e^{\alpha a^{\dagger} - \alpha^{*} a} \vert 0 \rangle \nonumber \\
  & = & e^{-\frac{|\alpha|^{2} }{2} } \sum_{n=0}^{\infty} \frac{\alpha^{n} }{ \sqrt{n \ !} } \vert n \rangle ,
\end{eqnarray}  
as proposed by Glauber\cite{glauber1,glauber2,glauber3}.

\subsection{Coherent State for the $\mathfrak{su}(2)$ Lie Algebra}

For the Lie group $SU(2)$ with Lie algebra $\mathfrak{su}(2)$, we have the operators $\left\lbrace J_{x}, J_{y}, J_{z} \right\rbrace$ that satisfy the commutation relations
\begin{eqnarray}
    \left[ J_{i},J_{j} \right] & = & i\epsilon_{ijk}J_{k},
\end{eqnarray}
where $\epsilon_{ijk}$ is the Levi-Civita symbols. Setting $J_{\pm} \ = \ J_{x} \pm iJ_{y}$ we have
\begin{eqnarray}
  \left[ J_{+},J_{-} \right] & = & 2J_{z}, \ \   \left[ J_{z},J_{\pm} \right] \ = \ \pm J_{\pm}.
\end{eqnarray}

The $SU(2)$ Lie group is compact, thus all irreducible representation is finite dimensional and it can be indexed by the symbol $j$. Then we can define the Dicke states
\begin{eqnarray}
  J^{2} \vert j,m \rangle & = & j(j+1) \vert j,m \rangle, \nonumber \\
  J_{z} \vert j,m \rangle & = & m  \vert j,m \rangle,
\end{eqnarray}
where $J^{2} \ = \ J_{x}^{2} + J_{y}^{2} + J_{z}^{2}$ is the Casimir operator.

For a reference state $ \vert j, -j \rangle$ the maximum stability subgroup is the $U(1)$ such that for the coset space $SU(2) / U(1)$ we have
  \begin{eqnarray}
  e^{\zeta J_{+}- \zeta^{*} J_{-}}  & \in &    SU(2) / U(1) , \ \ \zeta \in \mathbb{C},
  \end{eqnarray}
so that the Coherent State of $\mathfrak{su}(2)$ 
  \begin{eqnarray}
  \vert z \rangle & = & e^{\zeta J_{+}- \zeta^{*} J_{-}} \vert j,-j\rangle \nonumber \\
   & = &  \sum_{m=-j}^{j} \frac{z^{j+m} }{ (1+|z|^{2})^{j} } \sqrt{\binom{2j}{j+m}} \vert j,m\rangle,
\end{eqnarray}
with $\frac{z}{\sqrt{1+\vert z\vert^{2}}}  \ = \ \frac{\zeta \sin\vert \zeta \vert }{ \vert \zeta \vert  }$, as proposed by Atkins\cite{atkins} and Arecchi\cite{arecchi}.

\subsection{Coherent State for the $\mathfrak{su}(1,1)$ Lie Algebra}

Another case that we will analyze is the non-compact Lie group $SU(1,1)$, with Lie algebra $\mathfrak{su}(1,1)$ and generators $\left\lbrace K_{1}, K_{2}, K_{0}  \right\rbrace$ that satisfy the relations
\begin{eqnarray}
    \left[ K_{1},K_{2} \right] \ = \ -i K_{0}, &  &  \left[ K_{2},K_{0} \right] \ = \  i K_{1}, \nonumber \\
    \left[ K_{0},K_{1} \right] & = &  i K_{2}; \ \
\end{eqnarray}
also we have the relations 
\begin{eqnarray}
    \left[ K_{0},K_{\pm} \right] & = & \pm K_{\pm}, \ \ 
    \left[ K_{-},K_{+} \right] \ = \  2 K_{0}. 
\end{eqnarray}
with  $K_{\pm} \ = \ \pm i \left( K_{1} \pm i K_{2} \right)$.

Any irreducible representation is infinite dimensional, indexed by $k$ and $m$ so that 
\begin{eqnarray}
  K^{2} \vert k,m \rangle & = & k(k-1) \vert k,m \rangle, \nonumber \\
     K_{0} \vert k,m \rangle & = & (k+m)  \vert k,m \rangle,
\end{eqnarray}
where $K^{2} \ = \ K_{0}^{2} -K_{1}^{2} - K_{2}^{2} $ is the Casimir operator, $k = 1, \frac{3}{2}, 2, \frac{5}{2}, ...$ is the Bargmann index and $m \in \mathbb{N}$.

Taking the reference state as $\vert k,0 \rangle$ the maximum stability subgroup is again the $U(1)$. For the coset space $SU(1,1) / U(1)$ one has
  \begin{eqnarray}
  e^{\alpha K_{+}- \alpha^{*} K_{-}}  & \in &    SU(1,1) / U(1) , \ \ \alpha \in \mathbb{C}.
  \end{eqnarray}

Thus the Coherent state  is 
\begin{eqnarray}
 \vert \zeta,k \rangle & = & e^{\alpha K_{+} -\alpha^{*} K_{-} }  \vert k,0 \rangle \nonumber \\
          & = & (1-|\zeta|^{2})^{k} \sum_{m=0}^{+\infty} \sqrt{ \frac{\Gamma(2k+m)  }{m! \Gamma(2k) } } \zeta^{m} \vert k,m \rangle , \qquad
\end{eqnarray}
with $\zeta \ = \  \frac{\alpha}{|\alpha|} tanh |\alpha |$, as proposed by Barut and Girardello\cite{barut}, with $\Gamma$ be the Gamma Function.

\section{Thermofield Dynamics}\label{sec3}

Thermal effects in quantum theory were introduced in consistent way by $i)$ Matsubara\cite{matsubara} in 1955, known as  imaginary time formalism using the Wick rotation, $ii)$ Schwinger\cite{schwinger} and Keldsh\cite{keldsh} in the sixties with a real time formalism using the closed-time path formulation and $iii)$ Takahashi and Umezawa\cite{takahashi} in 1975 with the Thermofield Dynamics (TFD) formalism, which requires doubling of the Hilbert space.

In this paper we will explore the TFD formalism \cite{khanna} whose main propriety is the duplication of the original Hilbert space, preserving the structure of the operators algebra  and the commutation relations. The basic idea of this formalism is to look for a state $\vert 0 (\beta) \rangle$, namely thermal vacuum, such that the ensemble average of a operator is equal to the mean value, i.e., 
\begin{equation}
  \langle A \rangle := \langle 0(\beta) \vert A \vert 0(\beta) \rangle.
\end{equation}

If we assume that $\vert 0 (\beta) \rangle \in \mathcal{H}$, we can span this in terms of a Hamiltonian basis $\vert n \rangle$ resulting in $ \langle n \vert 0(\beta)\rangle \ = \ g_{n}(\beta)$. For the ensemble average  be equal to the mean value 
\begin{eqnarray}
  \langle 0(\beta) \vert A \vert 0(\beta) \rangle &=& \sum_{n,m} g^{*}_{n}(\beta) \langle n \vert A \vert m \rangle g_{m}(\beta) \nonumber \\
    &=&\sum_{n}\frac{ e^{-\beta E_{n} } \langle n \vert A \vert n \rangle }{Z(\beta)},
\end{eqnarray}
that imposes the condition on the coefficients $g_{m}(\beta)$ and $g^{*}_{n}(\beta)$
\begin{eqnarray}
  g^{*}_{n}(\beta)g_{m}(\beta) & = & \frac{1}{Z(\beta)} e^{-\beta E_{n}} \delta_{n,m}, \label{eqgngmbeta}
\end{eqnarray}
where $\delta_{n,m}$ is the Kronecker delta. The equation (\ref{eqgngmbeta}), like an orthogonality condition, cannot be satisfied by c-numbers, so $\vert 0 (\beta) \rangle$ cannot be an element of the original Hilbert space. One possibility explored by Takahashi and Umezawa\cite{takahashi} is by introducing a doubling of the Hilbert space $\mathcal{H} $, denoted by $\widetilde{\mathcal{H}}$, such that a vector basis is given by $ \vert n,\widetilde{m} \rangle \ \in \ \mathcal{H}\otimes \widetilde{\mathcal{H}}$. The idea of doubling the Hilbert space to introduce the thermal effect had already been proposed by Araki and Woods\cite{araki} in their works on Quantum Field Theory, so that doubled Hilbert space is not a single feature of TFD only.

In that case the resulting Thermal Vacuum 
$ \vert 0(\beta) \rangle  \in  \mathcal{H}\otimes \widetilde{\mathcal{H}}$, is
 \begin{eqnarray}
   \vert 0(\beta) \rangle  & = &  \sum_{n}\frac{e^{-\frac{\beta E_n}{2}}}{\sqrt{Z(\beta)}} \vert n,\widetilde{n} \rangle ,
\end{eqnarray}
and we can introduce a unitary transformation that maps the double vacuum $ \vert 0, \widetilde{0} \rangle$ into the thermal vacuum, namely Bogoliubov transformation $U(\beta)$
\begin{eqnarray}
 \vert 0(\beta) \rangle & = &  U(\beta)  \vert 0,\widetilde{0} \rangle.
\end{eqnarray}
In that way we can introduce a  notion of thermal operator as 
$  A(\beta) \ = \ U(\beta) \ A \ U^{\dagger}(\beta),$
where $  \beta \ = \ \frac{1 }{k_{b} T}$, $k_{b}$ is the Boltzmann constant and $T$ is the temperature.

\section{Generalized Thermal Coherent State}\label{sec4}

In order to derive the Generalized Thermal Coherent State in the TFD approach for multi-modes, let   $  G_{1} \times G_{2} \times \ldots \times G_{n}$  and $ \widetilde{G_{1}} \times \widetilde{G_{2}} \times \ldots \times \widetilde{G_{n}}$ be the product of arbitrary Lie groups with $ \Pi (g_{1})\otimes \Pi (g_{2}) \otimes \ldots \otimes \Pi (g_{n}) $ and $\widetilde{\Pi }(\widetilde{g_{1}}) \otimes \widetilde{\Pi }(\widetilde{g_{2}}) \otimes \ldots \otimes \widetilde{\Pi }(\widetilde{g_{n}}) $ two unitary irreducible representations of $  G_{1} \times G_{2} \times \ldots \times G_{n}$  and $ \widetilde{G_{1}} \times \widetilde{G_{2}} \times \ldots \times \widetilde{G_{n}}$, 
acting in the Hilbert space $\mathcal{H}=\mathcal{H}^{1}\otimes\mathcal{H}^{2}\otimes \ldots \otimes \mathcal{H}^{n} $ and $\widetilde{\mathcal{H}}=\widetilde{\mathcal{H}^{1}}\otimes\widetilde{\mathcal{H}^{2}}\otimes \ldots \otimes \widetilde{\mathcal{H}^{n}}$, respectively. 
Suppose $\left\lbrace (h_{1},  h_{2},  \ldots ,  h_{n} ) \right\rbrace \in H_{1}\times H_{2}\times \ldots \times H_{n} $ and $\left\lbrace (\widetilde{h}_{1},\widetilde{h}_{2},\ldots , \widetilde{h}_{n} ) \right\rbrace \in \widetilde{H_{1}}\times \widetilde{H_{2}}\times \ldots \times \widetilde{H_{n}}  $ are isotropy subgroups of $  G_{1} \times G_{2} \times \ldots \times G_{n}$  and $ \widetilde{G_{1}} \times \widetilde{G_{2}} \times \ldots \times \widetilde{G_{n}}$ for the states $\vert \psi_{0} \rangle $ $\left( \vert \psi_{0} \rangle \in \mathcal{H} \right) $ and $\vert \widetilde{\psi_{0} } \rangle $ $\left( \vert \widetilde{\psi_{0} } \rangle \in \widetilde{\mathcal{H}} \right) $. Their elements satisfy
\begin{eqnarray}
  \Pi (h_{1}) \Pi (h_{2})  \ldots  \Pi (h_{n}) \vert \psi_{0}  \rangle & = & \exp\left[  i\sum_{k=1}^{n} \phi_{k}(h_{k}) \right] \vert \psi_{0} \rangle , \nonumber \\
    \widetilde{\Pi }(\widetilde{h_{1}}) \widetilde{\Pi }(\widetilde{h_{2}})  \ldots  \widetilde{\Pi }(\widetilde{h_{n}}) \vert \widetilde{\psi}_{0}  \rangle & = & \exp\left[  i\sum_{k=1}^{n} \widetilde{\phi_{k}}(\widetilde{h}_{k}) \right] \vert \widetilde{\psi}_{0} \rangle ,
\end{eqnarray}
 with  $\exp[  i\phi_{i} (h_{i}) ] $ and $\exp[  i\widetilde{\phi_{i}} (\widetilde{h_{i}} ) ]$ phase factors.

For every elements $(g_{1}, g_{2}, \ldots , g_{n} )\in G_{1} \times G_{2} \times \ldots \times G_{n} $ and $(\widetilde{g}_{1},\widetilde{g}_{2},\ldots , \widetilde{g}_{n} ) \in \widetilde{G_{1}} \times \widetilde{G_{2}} \times \ldots \times \widetilde{G_{n}}$ we can obtain an unique decomposition
\begin{eqnarray}
  (g_{1}, g_{2}, \ldots , g_{n} ) & = & \Omega (h_{1},  h_{2},  \ldots ,  h_{n} ), \nonumber \\
 (\widetilde{g}_{1},\widetilde{g}_{2},\ldots , \widetilde{g}_{n} ) & = & \widetilde{\Omega} (\widetilde{h}_{1},\widetilde{h}_{2},\ldots , \widetilde{h}_{n} ),
\end{eqnarray}
with $\Omega \in (G_{1}\times G_{2}\times \ldots \times G_{n} )/(H_{1}\times H_{2}\times \ldots \times H_{n} )$ and $\widetilde{\Omega} \in ( \widetilde{G_{1}} \times \widetilde{G_{2}} \times \ldots \times \widetilde{G_{n}})/ ( \widetilde{H_{1}}\times \widetilde{H_{2}}\times \ldots \times \widetilde{H_{n}} ) $.

The action of an arbitrary element $(g_{1}, g_{2}, \ldots , g_{n} ) \times (\widetilde{g}_{1},\widetilde{g}_{2},\ldots , \widetilde{g}_{n} ) \in (G_{1}\times G_{2}\times \ldots \times G_{n} ) \times ( \widetilde{G_{1}} \times \widetilde{G_{2}} \times \ldots \times \widetilde{G_{n}}) $ on $\vert \psi_{0} \rangle \otimes \vert \widetilde{\psi}_{0} \rangle$ is given by
\begin{eqnarray}
\hspace{-2.0cm}  \Pi'(g \times \widetilde{g}) \vert \psi_{0}, \widetilde{\psi}_{0} \rangle & = & \Pi'(\Omega \times \widetilde{\Omega}) \Pi'((h_{1}, h_{2}, \ldots , h_{n} ) \times (\widetilde{h}_{1},\widetilde{h}_{2},\ldots , \widetilde{h}_{n} )) \vert \psi_{0}, \widetilde{\psi}_{0} \rangle \nonumber \\
\hspace{-2.0cm}  & = & \Pi(\Omega)\widetilde{\Pi}(\widetilde{\Omega})  \exp \left[ i\phi(h,\widetilde{h}) \right]  \vert \psi_{0}, \widetilde{\psi}_{0} \rangle, \qquad
\end{eqnarray}
where $\Pi' \equiv \Pi \otimes \widetilde{\Pi}$ is an unitary irreducible representation of $ (G_{1}\times G_{2}\times \ldots \times G_{n} ) \times ( \widetilde{G_{1}} \times \widetilde{G_{2}} \times \ldots \times \widetilde{G_{n}}) $  with phase factor  $\phi(h,\widetilde{h}) \equiv \sum_{k=1}^{n} \left[ \phi_{k}(h_{k}) +\widetilde{\phi_{k}}(\widetilde{h}_{k}) \right] $.

The double coherent states are then defined by
\begin{eqnarray}
  \vert \Lambda , \widetilde{\Lambda}, \Omega \times \widetilde{\Omega} \rangle & = & \Pi'(\Omega \times \widetilde{\Omega})  \vert \psi_{0}, \widetilde{\psi}_{0} \rangle \nonumber \\
  & = & \Pi(\Omega) \widetilde{\Pi} (\widetilde{\Omega})  \vert \psi_{0}, \widetilde{\psi}_{0} \rangle .
\end{eqnarray}

In other words, we consider that there is an one-to-one correspondence with the coset space
\begin{eqnarray}
  (G_{k} \times \widetilde{G}_{k})/ (H_{k} \times \widetilde{H}_{k}) & \simeq &  G_{k}/H_{k} \times \widetilde{G}_{k} / \widetilde{H}_{k} ,
\end{eqnarray} 
for $k\in\left\lbrace 1,2, \ldots , n \right\rbrace$. Thus we define the generalized thermal coherent state by
\begin{eqnarray}
  \vert \Lambda, \widetilde{\Lambda}, \Omega \times \widetilde{\Omega} ,\beta \rangle & = & U(\beta) \Pi(\Omega) \widetilde{\Pi}(\widetilde{\Omega})  \vert \psi_{0}, \widetilde{\psi}_{0} \rangle \nonumber \\
  & = & \Pi(\Omega,\beta)  \widetilde{\Pi}(\widetilde{\Omega},\beta)  \vert \psi_{0}, \widetilde{\psi}_{0} \rangle. \label{eq1234}
\end{eqnarray}

In (\ref{eq1234}) we use that $ \Pi(\Omega,\beta)  = U(\beta) \Pi (\Omega) U^{\dagger}(\beta)$ and $ \widetilde{\Pi}(\widetilde{\Omega},\beta) = U(\beta) \widetilde{\Pi}(\widetilde{\Omega}) U^{\dagger}(\beta) $, with $U(\beta)$ the Bogoliubov transformation \cite{khanna} that introduces the thermal effects.
 Two thermal states corresponding to same coset $\Omega \times \widetilde{\Omega}$ differ by a phase factor, i.e.
\begin{eqnarray}
  \vert \psi ( g_{1}, g_{2}, \ldots , g_{n} ), \widetilde{\psi}( \widetilde{g}_{1},\widetilde{g}_{2},\ldots , \widetilde{g}_{n}),\beta \rangle & = & \exp(i\alpha)  \vert \psi(g'_{1}, g'_{2}, \ldots , g'_{n}), \nonumber \\
  & & \widetilde{\psi}(\widetilde{g'}_{1},\widetilde{g'}_{2},\ldots , \widetilde{g'}_{n}), \beta \rangle, 
\end{eqnarray} 
where $g_{1}, g_{2}, \ldots , g_{n}=\Omega h_{1}, h_{2}, \ldots , h_{n}  \left( \widetilde{g}_{1},\widetilde{g}_{2},\ldots , \widetilde{g}_{n} = \widetilde{\Omega}\widetilde{h}_{1},\widetilde{h}_{2},\ldots , \widetilde{h}_{n} \right) $ and $g'_{1},  g'_{2}, $ $ \ldots , g'_{n} =\Omega h'_{1}, h'_{2}, \ldots , h'_{n} \ \left( \widetilde{g'}_{1},\widetilde{g'}_{2},\ldots , \widetilde{g'}_{n} = \widetilde{\Omega}\widetilde{h'}_{1},\widetilde{h'}_{2},\ldots , \widetilde{h'}_{n} \right) $.


Let $\mathfrak{g}_{1}\oplus \mathfrak{g}_{2}\oplus \ldots \oplus \mathfrak{g}_{n} $ and $\widetilde{\mathfrak{g}_{1}} \oplus \widetilde{\mathfrak{g}_{2}} \oplus \ldots \oplus \widetilde{\mathfrak{g}_{n}} $ be Lie algebras associated to Lie Groups $  G_{1} \times G_{2} \times \ldots \times G_{n}$  and $ \widetilde{G_{1}} \times \widetilde{G_{2}} \times \ldots \times \widetilde{G_{n}}$. If $\mathfrak{g}_{1}\oplus \mathfrak{g}_{2}\oplus \ldots \oplus \mathfrak{g}_{n} $ and $ \widetilde{\mathfrak{g}_{1}} \oplus \widetilde{\mathfrak{g}_{2}} \oplus \ldots \oplus \widetilde{\mathfrak{g}_{n}} $ are semi-simple algebras we have the Cartan basis $\left\lbrace H^{m}_{\alpha},E^{m}_{\alpha} \right\rbrace\in \mathfrak{g}_{m} $ given by
\begin{eqnarray}
  \left[ H_{i}^{m}, H_{j}^{m} \right] & = & 0, \qquad  \qquad \ \ \left[ \widetilde{H}_{i}^{m}, \widetilde{H}_{j}^{m} \right] \  = \  0, \nonumber \\
  \left[ H_{i}^{m}, E_{\alpha}^{m} \right] & = & \alpha_{i} E_{\alpha}^{m}, \qquad \ \ \ \left[ \widetilde{H}_{i}^{m}, \widetilde{E}_{\alpha}^{m} \right] \  = \ \widetilde{\alpha}_{i} \widetilde{E}_{\widetilde{\alpha} }^{m}, \nonumber \\
 \left[ E_{\alpha}^{m}, E_{-\alpha}^{m} \right] & = & \alpha_{i} H_{i}^{m}, \ \ \ \ \ \ \ \ \left[ \widetilde{E}_{\alpha}^{m}, \widetilde{E}_{-\alpha}^{m} \right] \  = \ \widetilde{\alpha}_{i} \widetilde{H}_{ i }^{m}, \nonumber \\
 \left[ E_{\alpha}^{m}, E_{\varepsilon}^{m} \right] & = & N_{\alpha,\varepsilon}^{m} E_{\alpha+\varepsilon}^{m}, \ \ \ \ \ \left[ \widetilde{E}_{\alpha}^{m}, \widetilde{E}_{\varepsilon }^{m} \right] \  = \ \widetilde{N}_{\widetilde{\alpha}^{m},   \widetilde{\varepsilon } }    \widetilde{E}_{ \widetilde{\alpha}+\widetilde{\varepsilon } }^{m} ,  \qquad
\end{eqnarray}
following the standard notation \cite{bookliealgebra}, for different Lie algebra all commutation relations are null.

It follows that we have a Lie algebra $\mathfrak{g}^{T}$ associated to group $ \left( G_{1} \times G_{2} \times \right. $ $\left. \ldots \times G_{n} \right)  \times \left( \widetilde{G_{1}} \times \widetilde{G_{2}} \times \ldots \times \widetilde{G_{n}} \right) $ given by $\mathfrak{g}^{T} = \left( \mathfrak{g}_{1}\oplus \mathfrak{g}_{2}\oplus \ldots \oplus \mathfrak{g}_{n} \right) \oplus \left( \widetilde{\mathfrak{g}_{1}} \oplus \widetilde{\mathfrak{g}_{2}} \oplus \ldots \oplus \widetilde{\mathfrak{g}_{n}} \right) $. For the semi-simple Lie algebra $\mathfrak{g}^{T} $, the Cartan basis is given by elements
\begin{eqnarray} 
  \left( H_{i}^{m},\widetilde{H}_{j}^{n} \right), \ \left( H_{i}^{m}, \widetilde{E}_{\widetilde{\alpha} }^{n} \right), & ... & \left( E_{\alpha}^{m}, \widetilde{E}_{\varepsilon }^{n} \right),
\end{eqnarray}
and the Lie bracket is defined as
\begin{eqnarray}
  \left[ \left( x_{i}, \widetilde{x}_{i} \right), \left( y_{i}, \widetilde{y}_{i} \right) \right] & = &   \left( \left[ x_{i}, y_{i} \right], \left[ \widetilde{x}_{i}, \widetilde{y}_{i} \right] \right),
\end{eqnarray}
with $x_{i},y_{i} \in \mathfrak{g}_{1}\oplus \mathfrak{g}_{2}\oplus \ldots \oplus \mathfrak{g}_{n}  $ and $\widetilde{x}_{i}, \widetilde{y}_{i} \in  \widetilde{\mathfrak{g}_{1}} \oplus \widetilde{\mathfrak{g}_{2}} \oplus \ldots \oplus \widetilde{\mathfrak{g}_{n}}$.
 If we define

\begin{eqnarray}
  \Pi\left( E_{\alpha}^{m},\beta \right) & = & U(\beta) \Pi\left( E_{\alpha}^{m}  \right) U^{\dagger} (\beta) , \nonumber \\
 \Pi\left( E_{-\alpha}^{m},\beta \right) & = & U(\beta) \Pi\left( E_{-\alpha}^{m} \right) U^{\dagger} (\beta) , \nonumber \\
  \widetilde{\Pi}\left( \widetilde{E}_{\widetilde{\alpha} }^{m},\beta \right) & = & U(\beta) \widetilde{\Pi}\left( \widetilde{E}_{\widetilde{\alpha} }^{m} \right) U^{\dagger} (\beta) , \nonumber \\
  \widetilde{\Pi}\left( \widetilde{E}_{-\widetilde{\alpha}}^{m},\beta \right) & = & U(\beta) \widetilde{\Pi}\left( \widetilde{E}_{-\widetilde{\alpha} }^{m} \right) U^{\dagger} (\beta) ,
\end{eqnarray}
we can rewrite the generalized thermal coherent state as
\begin{eqnarray}
\hspace{-2.1cm}  \vert \Lambda, \widetilde{\Lambda}, \Omega \times \widetilde{\Omega}, \beta \rangle & = &
  \exp\left\lbrace \sum_{\alpha,m} \left[ \eta_{\alpha,m} \Pi\left( E_{\alpha}^{m},\beta \right)  - \eta^{*}_{\alpha,m}  \Pi\left(E_{-\alpha}^{m},\beta \right) \right] \right\rbrace \nonumber \\
\hspace{-2.1cm}  & & \times \exp\left\lbrace \sum_{\widetilde{\alpha},m } \left[ \widetilde{\eta}_{ \widetilde{\alpha},m }       \widetilde{\Pi}\left(\widetilde{E}_{\widetilde{\alpha} }^{m},\beta \right) - \widetilde{\eta}^{*}_{\widetilde{\alpha},m} \widetilde{\Pi}\left(\widetilde{E}_{-\widetilde{\alpha} }^{m},\beta \right) \right] \right\rbrace  \vert \psi_{0},\widetilde{\psi}_{0}, \beta \rangle . \qquad
\end{eqnarray}

According to the previous formulation and the references \cite{perelomov86,zhang90} it is easy to show the following properties:

\begin{itemize}
  \item[$i)$]  Non-orthogonality
  \begin{eqnarray}
    \langle \beta ; \Omega\times \widetilde{\Omega}, \widetilde{\Lambda},\Lambda \vert \beta; \Lambda, \widetilde{\Lambda}, \Omega' \times \widetilde{\Omega}' \rangle 
    & \neq & 0,
  \end{eqnarray}
for $\Omega \neq \Omega' \in G/H$ and $\widetilde{\Omega} \neq \widetilde{\Omega}' \in \widetilde{G}/ \widetilde{H}$, being, however, normalized 
  \begin{eqnarray}
    \langle \beta ; \Omega\times \widetilde{\Omega}, \widetilde{\Lambda},\Lambda  \vert \beta; \Lambda, \widetilde{\Lambda}, \Omega \times \widetilde{\Omega} \rangle & = &  \langle \beta ; \widetilde{\psi}_{0}, \psi_{0} \vert \Pi \left[ (\Omega \times \widetilde{\Omega})^{-1},\beta \right] \nonumber \\
   & &  \times  \Pi \left[ \Omega\times\widetilde{\Omega},\beta \right]\vert \beta ; \psi_{0}, \widetilde{\psi}_{0} \rangle \nonumber \\
 & = &   
1.
  \end{eqnarray}

\item[$ii)$] Over-completeness
%
\begin{eqnarray}
\hspace{-0.7cm} \int d\mu (\Omega \times \widetilde{\Omega}, \beta)  \  \vert \beta; \Lambda, \widetilde{\Lambda}, \Omega \times \widetilde{\Omega} \rangle \langle  \beta; \Omega \times \widetilde{\Omega}, \widetilde{\Lambda}, \Lambda  \vert & = & I \nonumber \\ \label{cap5323454}
\end{eqnarray}
so any thermal state can be expand in terms of the Thermal Coherent State, i.e.
\begin{eqnarray}
  \vert \psi(\beta) \rangle & = & \int d\mu (\Omega \times \widetilde{\Omega},\beta)   f_{\Lambda} (\Omega \times \widetilde{\Omega},\beta) \nonumber \\
  & & \times N^{-1/2}(\Omega \times \widetilde{\Omega},\beta) \vert \beta; \Lambda, \widetilde{\Lambda}, \Omega \times \widetilde{\Omega} \rangle, \ \qquad
\end{eqnarray}
where  $f_{\Lambda} (\Omega \times \widetilde{\Omega},\beta)$ is the coefficient of the state defined over  $ (G\times \widetilde{G})/ (H\times \widetilde{H})$ and $N(\Omega\times \widetilde{\Omega},\beta)$ is the normalization constant.

\end{itemize}

\section{Applications}\label{sec5}

 In this section we apply our formulation to obtain the generalized thermal coherent state for $\mathfrak{su}(2)$ and $\mathfrak{su}(1,1)$ Lie algebras. We have obtained their thermal density operator, with was used to calculated the thermal quantum Fidelity and the thermal Wigner function.

\subsection{Thermal Coherent State of $\mathfrak{su}(2)$}

Consider now atomic coherent states, also known as spin coherent states \cite{zhang90,arecchi}. These states can be realized by Bose-Einstein condensates and applied in the analysis of entanglement in Quantum Information\cite{bigelow,gross12}. In this case, we have the representative coset given by  $U(\beta) \left( \frac{SU(2) \times SU(2) }{ U(1) \times U(1) } \right) U^{\dagger}(\beta) $, so that
\begin{eqnarray}
  \vert z,\widetilde{z}, \Omega \times \widetilde{\Omega},\beta \rangle & = & \exp\left[ \tau J_{+}(\beta) - \tau^{*} J_{-}(\beta) \right] \nonumber \\
  & & \times  \exp\left[ \gamma \widetilde{J}_{+}(\beta) - \gamma^{*} \widetilde{J}_{-}(\beta) \right]  \vert j,-j;\widetilde{j},\widetilde{-j};\beta \rangle,
\end{eqnarray}
with the commutation relations
\begin{eqnarray}
  \left[ J_{+}(\beta), J_{-}(\beta) \right] & = & 2J_{z} (\beta), \qquad \qquad 
  \left[ \widetilde{J}_{+}(\beta), \widetilde{J}_{-}(\beta) \right] \ = \ 2\widetilde{J}_{z} (\beta), \nonumber \\
  \left[ J_{z}(\beta), J_{\pm}(\beta) \right] & = & \pm J_{\pm} (\beta), \qquad  \ \ \ \ \
  \left[ \widetilde{J}_{z}(\beta), \widetilde{J}_{\pm}(\beta) \right] \ = \ \pm \widetilde{J}_{\pm}(\beta) .
\end{eqnarray}
A thermal coherent state  of the harmonic oscillator can be built  considering the coset $ U(\beta) \left( \frac{ \displaystyle W_{1} \times W_{1}  }{U(1)\times U(1)} \right) U^{\dagger}(\beta)$, with the resulting state\cite{khanna}
\begin{eqnarray}
  \vert \alpha(\beta) \rangle & = & U(\beta) \exp\left[  \alpha a^{\dagger} - \alpha^{*} a \right] U^{\dagger}(\beta) \vert 0(\beta) \rangle, \ \ \ 
\end{eqnarray}
where $W_{1}$ is the Weyl algebra; this procedure is important because the  coherent  state $\vert \alpha(\beta) \rangle$ reduces to the pure state $\vert \alpha \rangle$ in  the limit $T \rightarrow 0$ ( $T$ is a temperature) or $\beta \rightarrow + \infty$.    
The limit of the temperature going to zero can become quite problematic to perform in situation like phase transition, so our interest are in cases that the system is in a single phase.
 In according to this scheme we propose the state
\begin{eqnarray}
  \vert z (\beta) \rangle & = &    \exp\left[ \eta J_{+}(\beta) - \eta^{*} J_{-}(\beta) \right]  \vert \beta; j,-j,\widetilde{0}, \widetilde{0} \rangle \nonumber \\
  & = & \frac{\exp\left[  zJ_{+}(\beta) \right] }{\left( 1+\vert z\vert^{2}  \right)^{j} } U(\beta) \vert j,-j,\widetilde{0}, \widetilde{0} \rangle,
\end{eqnarray}
where $  \frac{z}{\sqrt{1+\vert z\vert^{2}}}  \ = \ \frac{\eta \sin\vert \eta \vert }{ \vert \eta \vert  }$ and
Baker-Campbell-Hausdorff formula was used \cite{zhang90}.

Moreover, by using the two-boson Schwinger representation, we have
\begin{eqnarray}
  J_{+}(\beta) \ = \ a^{\dagger}_{1} (\beta) a_{2} (\beta), & &
  J_{-}(\beta) \ = \ a^{\dagger}_{2} (\beta) a_{1} (\beta) \nonumber \\
  J_{z}(\beta) \ = \ \frac{1}{2}\left[a^{\dagger}_{1} (\beta) a_{1} (\beta) \right. & - & \left. a^{\dagger}_{2} (\beta) a_{2} (\beta) \right],
\end{eqnarray}
and
\begin{eqnarray}
\hspace{-2.0cm}  \vert z(\beta) \rangle & = & \left( 1 + \vert z \vert^{2} \right)^{-j}  \sum_{m=-j}^{j}
  \sqrt{ \left(\begin{array}{c}
    2j \\ j+m
  \end{array}\right) }  \frac{ \left( a_{1}^{\dagger}(\beta) \right)^{j+m} \left( a_{2}^{\dagger}(\beta)
   \right)^{j-m}    }{ \sqrt{(j+m)!}  \sqrt{(j-m)!} }  z^{j+m}
   \vert 0(\beta) \rangle, \qquad \label{eq24}
\end{eqnarray}
where the Bogoliubov transformation is given by $U(\beta) \ = \ \exp\left[ -iG(\beta) \right]$, with
\begin{eqnarray}
  G(\beta) & = & \sum_{i=1}^{2}-i\theta_{i}(\beta) \left( \widetilde{a}_{i}a_{i} - \widetilde{a}_{i}^{\dagger} a_{i}^{\dagger}  \right). 
\end{eqnarray}

For the state given by (\ref{eq24}) we have the following properties
 
\begin{itemize}
  \item[$i)$] Non-orthogonality
  \begin{eqnarray}
    \langle z_{1} (\beta) \vert z_{2} (\beta) \rangle & = & \frac{\left(  1+z_{1}^{*}z_{2}\right)^{2j} }{ \left( 1 + \vert z_{1}\vert^{2} \right)^{j}  \left( 1 + \vert z_{2}\vert^{2} \right)^{j}  }. \qquad
  \end{eqnarray}

  \item[$ii)$]  Over-completeness
\begin{eqnarray}
\int d\mu \left( z(\beta),z^{*}(\beta) \right)  \vert z(\beta) \rangle \langle z(\beta) \vert & = & 1,
\end{eqnarray}
with
\begin{eqnarray}
  d\mu \left( z(\beta),z^{*}(\beta) \right) & = & \frac{2j+1}{\pi} \frac{ dz(\beta) dz^{*}(\beta) }{ \left( 1 + \vert z \vert^{2} \right)^{2} }. \qquad
\end{eqnarray}

\end{itemize}

Using the thermal average
\begin{eqnarray}
  \langle z(\beta) \vert \widehat{O} \vert z(\beta) \rangle & = & Tr \rho_{\vert z(\beta) \rangle } \widehat{O},
\end{eqnarray}
we determine that the density operator for the thermal coherent state of $\mathfrak{su}(2)$ is 
\begin{eqnarray}
\rho_{\vert z(\beta) \rangle}  & = &  \sum_{m,m' \ = \ -j  }^{j}  \
\sum_{ n_{1},n_{2} \ = \ 0 }^{ + \infty}  C_{\stackrel{m,m'}{n_{1}n_{2}} }(z,\beta) \nonumber \\
 & & \times \vert n_{1} +j+m,n_{2}+j-m \rangle  \langle n_{1}+j+m',n_{2}+j-m'\vert, \ \ \ \label{rhothermsu2}
\end{eqnarray}
with
\begin{eqnarray}
\hspace{-2.0cm}  C_{\stackrel{m,m'}{n_{1}n_{2}} }(z,\beta)
   & = & \sqrt{ \left(\begin{array}{c}
     2j \\ j+m
   \end{array}\right)  }  \sqrt{ \left(\begin{array}{c}
     2j \\ j+m'
   \end{array}\right)  }    \frac{z^{j+m}(z^{*})^{j+m'}  \left[ \exp(-\beta\omega) \right]^{n_{1}+n_{2}} }{   (1+|z|^{2})^{2j} \sqrt{(j+m)!}  } \nonumber \\
 \hspace{-2.0cm}  & & \times \frac{  \left[  1-\exp(-\beta\omega) \right]^{2j+2}  }{\sqrt{(j-m)!}  \sqrt{(j+m')!}\sqrt{(j-m')!}  }  \sqrt{\frac{(n_{1}+j+m)!}{n_{1}!}} \sqrt{\frac{(n_{2}+j-m)!}{n_{2}!}} \nonumber \\
 \hspace{-2.0cm}  & & \times \sqrt{\frac{(n_{1}+j+m')!}{n_{1}!}} \sqrt{\frac{(n_{2}+j-m')!}{n_{2}!}} . \qquad
\end{eqnarray}

Eq. (\ref{rhothermsu2}) is the density matrix associated to state $\vert z(\beta) \rangle$. In the limit $ T \rightarrow 0 \ \left( \beta \rightarrow + \infty \right) $ we have recovered the state $\vert z \rangle$ \cite{zhang90}.

\subsubsection{$\mathfrak{su}(2)$ Thermal Fidelity \\ }

The Fidelity $F$ is a measure of distance in the Hilbert space that plays an important role in Quantum Information\cite{chuang}; $F \ \in \ \left[ 0,1 \right]$ is given by
\begin{eqnarray}
  F & = & \sqrt{\langle z \vert \rho_{ \vert z(\beta) \rangle} \vert z \rangle }, \label{fidelity01}
\end{eqnarray}
providing the distance between the $\mathfrak{su}(2)$ Thermal Coherent State and the usual non-thermal $\mathfrak{su}(2)$ Coherent State. For calculate the Fidelity we will use the equation (\ref{rhothermsu2}) in the expression (\ref{fidelity01}) of the Fidelity, that results in
\begin{eqnarray} \label{f0}
    F & = &  (1-e^{-\beta\omega})^{j+1}.
\end{eqnarray}

For $T \rightarrow 0$ we have $F \rightarrow 1$; so our $\mathfrak{su}(2)$ Thermal Coherent State coincides with the usual non-thermal coherent state. An increase of temperature in the thermal state results in a growth of distance in relation to the non thermal state. 


\subsubsection{Thermal Wigner Function \\ }

The Wigner function is a quasi-probability distribution whose  negative values are associated to the degree of non-classicality of the system\cite{khanna}. It is defined by
 \begin{eqnarray} 
\hspace{-2.5cm} f_{w}
 &=&\int_{-\infty}^{+\infty}dv_{1}e^{ip_{1}v_{1}}
  \int_{-\infty}^{+\infty}dv_{2}e^{ip_{2}v_{2}}  \langle q_1-\frac{v_1}{2},q_2-\frac{v_2}{2} \vert \rho_{ \vert z(\beta)\rangle} \vert q_1+\frac{v_1}{2},q_2+\frac{v_2}{2} \rangle. \
\end{eqnarray}

Using eq. (\ref{rhothermsu2}) that carries all information about the $\mathfrak{su}(2)$ Thermal Coherent State, we can find the expression of the Thermal Wigner Function as
 \begin{eqnarray}
\hspace{-3.1cm}  f_{w}(x_1,x_2;z,\beta)&=&
 \sum_{\stackrel{m,m'=-j}{n_{1},n_{2}=0}}^{j,\infty}
 \sqrt{\binom{2j}{j+m}}\sqrt{\binom{2j}{j+m'}} \frac{z^{j+m}(z^{*})^{j+m'} 
 \left( e^{-\beta\omega}\right)^{n_{1}+n_{2}}  \left( 1- e^{-\beta\omega} \right)^{2j+2}  }{
   (1+|z|^{2})^{2j}  } \nonumber \\ 
\hspace{-3.1cm}   & & \times \frac{4 min(n_{1}+j+m,n_{1}+j+m')! min(n_{2}+j-m,n_{2}+j-m')! }{  
  \sqrt { (j+m)!(j-m)!  (j+m')!(j-m')! } n_{1}! n_{2}! } \nonumber \\
\hspace{-3.1cm} & &\times e^{-x^{2}_{1} -x^{2}_{2}}  (-1)^{n_{1}+n_{2}+2j }  \frac{2^{max(n_{1}+j+m,n_{1}+j+m')}2^{max(n_{2}+j-m,n_{2}+j-m')} }{ 2^{n_{1}+n_{2}+2j} }  \nonumber \\
 \hspace{-3.1cm} & & \times \chi_{max(n_{1}+j+m,n_{1}+j+m')}^{|m-m'|} \chi_{max(n_{2}+j-m,n_{2}+j-m')}^{|m-m'|} \nonumber \\
 \hspace{-3.1cm} & & \times L_{min(n_{1}+j+m,n_{1}+j+m')}^{|m-m'|} (2x^{2}_{1} )
  L_{min(n_{2}+j-m,n_{2}+j-m')}^{|m-m'|} (2x^{2}_{2}) .
\end{eqnarray}
%
where $L_{n}^{\alpha}$ are associated Laguerre polynomials, $x_{\iota }=i \frac{p_{\iota}}{\sqrt{\omega}} +q_{\iota}\sqrt{\omega}$ with $\iota = \left\lbrace 1,2 \right\rbrace$ and 
\begin{equation}
  \left\lbrace\begin{array}{l}
 \chi_{i}^{|m-m'|} \chi_{j}^{|m-m'|} \ = \
(-x_{2}x_{1}^{*})^{|m-m'|}, \ \ \ 
\textrm{if} \ m<m' \ \textrm{or} \\    
\chi_{i}^{|m-m'|} \chi_{j}^{|m-m'|} \ = \
(-x^{*}_{2}x_{1})^{|m-m'|}, \ \ \ \textrm{if} \ m \geq m'.
\end{array} \right.  \label{fwignerxi23}
\end{equation} 
As example, we plot the thermal Wigner function of the coherent state of $\mathfrak{su}(2)$ in Fig. 1 for a temperature of $0.005K$.

\begin{figure}[!ht]  
\centering
\includegraphics[scale=0.075]{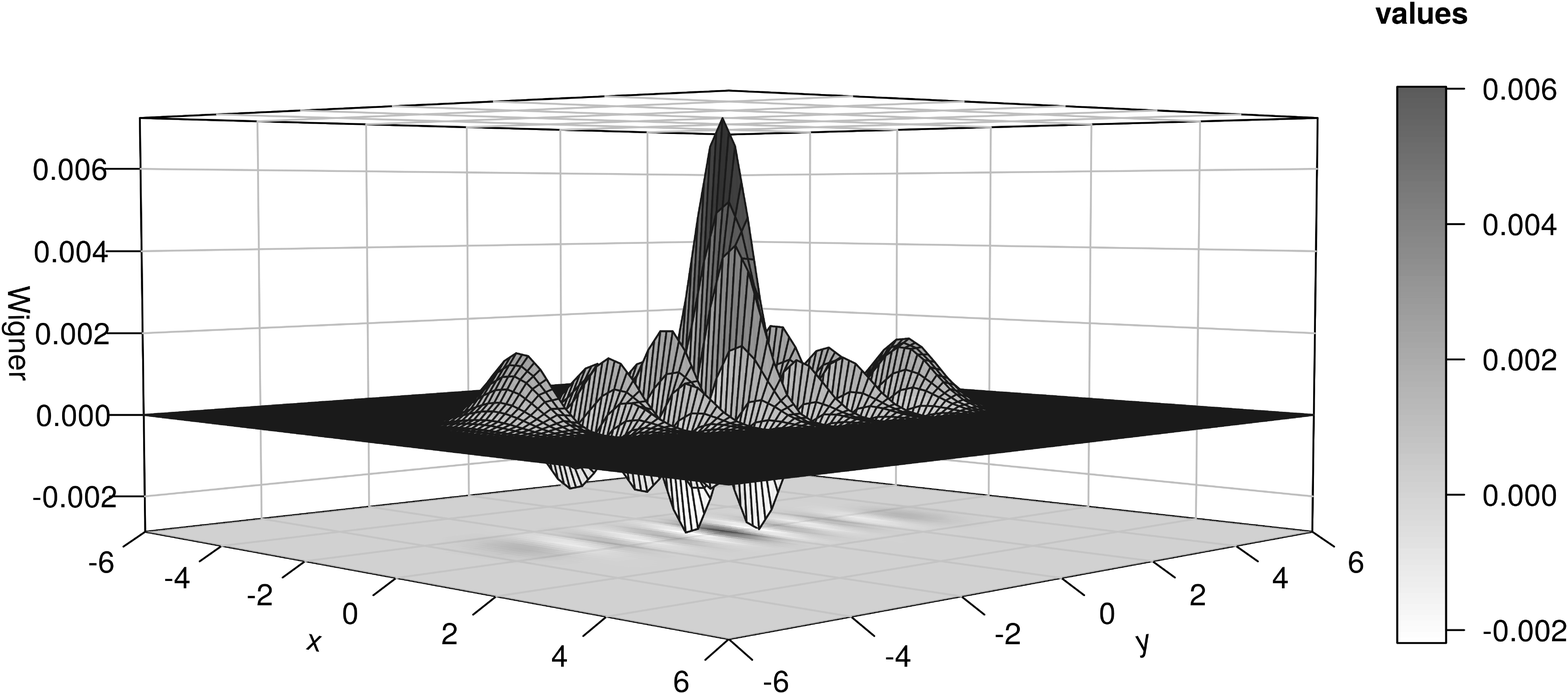}  \label{figura1}  
\caption{ Thermal Wigner function for $\mathfrak{su}(2)$ Lie algebra with $j=3$, $z=0.1$, $\omega=10^{7}Hz$ and $T=0.005K$.}  
\end{figure}

\subsection{Thermal Coherent State of $\mathfrak{su}(1,1)$}

Now we consider the case associated to the Lie algebra $\mathfrak{su}(1,1)$. These  states can be generated by Quantum Optics\cite{gerry85,wodkiewicz85} and play an important role in Quantum Metrology \cite{berrada13}. The representative coset is given by  $\displaystyle U(\beta) \left( \frac{SU(1,1) \times SU(1,1)}{U(1) \times U(1)}  \right) U^{\dagger}(\beta)$. Thus
\begin{eqnarray}
  \vert \zeta,\widetilde{\zeta},\Omega \times \widetilde{\Omega}, \beta \rangle & = & \exp\left[ \alpha K_{+}(\beta) - \alpha^{*} K_{-}(\beta)  \right] \nonumber \\
  & & \times \exp\left[ \sigma \widetilde{K}_{+}(\beta) - \sigma^{*} \widetilde{K}_{-}(\beta)  \right]  \vert k,0,\widetilde{k},0 ; \beta \rangle,
\end{eqnarray}
with the  commutation relations
\begin{eqnarray}
  \left[ K_{-}(\beta), K_{+}(\beta) \right] & = & 2K_{0} (\beta),  \qquad \qquad
  \left[ \widetilde{K}_{-}(\beta), \widetilde{K}_{+}(\beta) \right] \ = \ 2\widetilde{K}_{0} (\beta), \nonumber \\
  \left[ K_{0}(\beta), K_{\pm}(\beta) \right] & = & \pm K_{\pm} (\beta),  \qquad \ \ \ \ \ 
  \left[ \widetilde{K}_{0}(\beta), \widetilde{K}_{\pm}(\beta) \right] \ = \ \pm \widetilde{K}_{\pm}(\beta) .
\end{eqnarray}

Similarly to previous case, in order to ensure that for $T \rightarrow 0$ ($\beta \rightarrow +\infty$) the original state is preserved, we propose
\begin{eqnarray}
  \vert \zeta (\beta)  \rangle & = &   \exp\left[ \alpha K_{+}(\beta) - \alpha^{*} K_{-}(\beta) \right]  \vert \beta ; k,0,\widetilde{\frac{1}{2}}, \widetilde{0} \rangle \nonumber \\
  & = & \left( 1- |\zeta|^{2}  \right)^{k}   \exp\left[  \zeta K_{+}(\beta) \right]   \vert \beta ; k,0,\widetilde{\frac{1}{2} }, \widetilde{0} \rangle, \qquad
\end{eqnarray}
where we have used the Baker-Campbell-Hausdorff formula, denoted  $\zeta  =  e^{i\phi} \tanh ( r)$ and $\alpha \ = \ r e^{i\phi}$.
Using the two-boson representation
\begin{eqnarray}
  K_{+}(\beta) \ = \ a^{\dagger}_{1} (\beta) a^{\dagger}_{2} (\beta), & &  
  K_{-}(\beta) \ = \ a_{1} (\beta) a_{2} (\beta) \nonumber \\
  K_{0}(\beta) \ = \ \frac{1}{2}\left[a^{\dagger}_{1} (\beta) a_{1} (\beta) \right. &+& \left. a^{\dagger}_{2} (\beta) a_{2} (\beta) + 1 \right],
\end{eqnarray}
the correspondence $\vert k,m \rangle \mapsto \vert n+q,n \rangle $, $k = \frac{1}{2}(1+q)$ and $m=n$, we obtain
\begin{eqnarray}
  \vert \zeta(\beta) \rangle & = & \left( 1 - \vert \zeta \vert^{2} \right)^{\frac{1}{2}(1+q)}  \sum_{n=0}^{+\infty}
  \sqrt{ \frac{(q+n)!}{n!q!} }  \zeta^{n} 
 \vert n+q,n,\widetilde{0},\widetilde{0};\beta \rangle .
\end{eqnarray}

For states $\vert \zeta(\beta) \rangle$ the following properties are verified:

\begin{itemize}
  \item[$i)$] Non-orthogonality
\end{itemize}
  \begin{eqnarray}
\hspace{-1.0cm}     \langle \zeta_{1} (\beta) \vert \zeta_{2} (\beta) \rangle & = & \left( 1 - \vert \zeta_{1} \vert^{2} \right)^{\frac{1}{2}(1+q) } \left( 1 - \vert \zeta_{2} \vert^{2} \right)^{\frac{1}{2}(1+q) }  \left( 1 -  \zeta_{1}^{*} \zeta_{2} \right)^{-(1+q) }
  \end{eqnarray}
and
\begin{itemize}
  \item[$ii)$]  Over-completeness
\end{itemize} 
\begin{eqnarray}
\int d \mu \left( \zeta (\beta),\zeta^{*}(\beta) \right)  \vert \zeta(\beta) \rangle \langle \zeta(\beta) \vert & = & 1,
\end{eqnarray}
with

\begin{eqnarray}
  d \mu \left( \zeta(\beta),\zeta^{*}(\beta) \right) & = & \frac{2k-1}{\pi} \frac{ d\zeta(\beta) d\zeta^{*}(\beta) }{ \left( 1 - \vert \zeta \vert^{2} \right)^{2} }.
\end{eqnarray}

From the thermal average we obtain that the associated density operator is
\begin{eqnarray}
\rho_{\vert \zeta(\beta) \rangle}  & = &  \sum_{n,\bar{n} \ = \ 0  }^{+\infty}  \
\sum_{ n_{1},n_{2} \ = \ 0 }^{+\infty}  \Gamma_{ \stackrel{ n, \bar{n} }{  n_{1},  n_{2} }  } (\zeta,\beta) \nonumber \\
 & & \times \vert n_{1} + n + q, n_{2} + n \rangle \langle n_{1}+\bar{n}+q,n_{2}+\bar{n} \vert, \nonumber \\       \label{thermalrhosu11}
\end{eqnarray}
with
\begin{eqnarray}
  \Gamma_{ \stackrel{ n, \bar{n}}{ n_{1}, n_{2} } } (\zeta,\beta)
   & = & 
     \frac{  (1- \vert \zeta \vert^{2} )^{1+q} (\zeta^{*})^{\bar{n}} \zeta^{n} \left[ \exp(-\beta\omega) \right]^{n_{1}+n_{2}}   }{
   q! \ n! \ \bar{n}! } \nonumber \\
   & & \times  \left[  1-\exp(-\beta\omega) \right]^{n+\bar{n}+q+2}  \sqrt{\frac{(n_{2}+n)!}{n_{2}!}} \nonumber \\
   & & \times  \sqrt{\frac{(n_{1}+n+q)!}{n_{1}!}}  \sqrt{\frac{(n_{2}+\bar{n})!}{n_{2}!}}  \sqrt{\frac{(n_{1}+\bar{n}+q)!}{n_{1}!}}.
\end{eqnarray}

\subsubsection{$\mathfrak{su}(1,1)$ Thermal Fidelity \\ }

Similar to the previous section we can study the Fidelity of these states, with the intention of compare our $\mathfrak{su}(1,1)$ Thermal Coherent States with the non-thermal coherent states. In this case, the quantum Fidelity is
\begin{eqnarray}
   F & = & \sqrt{\langle \zeta \vert \rho_{\vert \zeta (\beta)\rangle} \vert \zeta \rangle }. \label{fidelity02}
\end{eqnarray}

Using eq. (\ref{thermalrhosu11}) we obtain 
\begin{eqnarray}
 \hspace{-1.8cm} F & = & \sum_{ n,\bar{n},n_{1} \ = \ 0 }^{+\infty}      \frac{  (1- \vert \zeta \vert^{2} )^{2+2q} \left( \vert \zeta \vert^{2} \right)^{n+\bar{n}+n_{1}}  \left[ \exp(-\beta\omega) \right]^{2n_{1}}  \left[  1-\exp(-\beta\omega) \right]^{n+\bar{n}+q+2}  }{
   \left( q!\right)^{2} \left( n_{1}! \right)^{2} \ n! \ \bar{n}! } \nonumber \\
 \hspace{-1.8cm}  & & \times  (n+n_{1}+q)!  (\bar{n}+n_{1}+q)! \ .
\end{eqnarray}
 
For $T \rightarrow 0$ the Fidelity go to $ F \rightarrow 1$, showing that for zero temperature we recover the usual state. For $T > 0$ the Fidelity is lower that $1$ evidencing that the $\mathfrak{su}(1,1)$ Thermal Coherent State is a new state differing from the usual case.

\subsubsection{Thermal Wigner Function \\ }

For $\mathfrak{su}(1,1)$ Thermal Coherent States, the Wigner function is given by
 \begin{eqnarray} \hspace{-1.0cm}
\hspace{-1.0cm} f_{w}
 &=&\int_{-\infty}^{+\infty}dv_{1}e^{\frac{ip_{1}v_{1}}{\hslash }}
  \int_{-\infty}^{+\infty}dv_{2}e^{\frac{ip_{2}v_{2}}{\hslash }}  \langle q_1-\frac{v_1}{2},q_2-\frac{v_2}{2} \vert \rho_{ \vert \zeta (\beta)\rangle} \vert q_1+\frac{v_1}{2},q_2+\frac{v_2}{2} \rangle 
\end{eqnarray} 
with $\rho_{ \vert \zeta (\beta)\rangle}$ given by equation (\ref{thermalrhosu11}). So we can obtain that
 \begin{eqnarray}
\hspace{-1.8cm} f_{w}
 &=&
 \sum_{n,\bar{n},n_{1},n_{2}=0  }^{+\infty}
      \frac{  (1- \vert \zeta \vert^{2} )^{1+q} (\zeta^{*})^{\bar{n}} \zeta^{n}  \left[  1-\exp(-\beta\omega) \right]^{n+\bar{n}+q+2}   \left[ \exp(-\beta\omega) \right]^{n_{1}+n_{2}} }{
   q! \ n! \ \bar{n}! } \nonumber \\
   & & \times  \sqrt{\frac{(n_{1}+n+q)!}{n_{1}!}} \sqrt{\frac{(n_{2}+n)!}{n_{2}!}}  \sqrt{\frac{(n_{1}+\bar{n}+q)!}{n_{1}!}} \sqrt{\frac{(n_{2}+\bar{n})!}{n_{2}!}}   \nonumber \\
\hspace{-1.8cm}   & & \times \frac{4 min(n_{1}+n+q,n_{1}+\bar{n}+q)! min(n_{2}+n,n_{2}+\bar{n})! e^{-x^{2}_{1} -x^{2}_{2}}  }{  
  \sqrt { (n_{1}+n+q)!(n_{1}+\bar{n}+q)!  (n_{2}+n)!(n_{2}+\bar{n})! }  } \nonumber \\
\hspace{-1.8cm} & &\times (-1)^{n_{1}+n_{2}+q }  \frac{2^{max(n_{1}+n+q,n_{1}+\bar{n}+q)}2^{max(n_{2}+n,n_{2}+\bar{n})} }{ 2^{n_{1}+n_{2}+q+n+\bar{n}} } \chi_{max(n_{1}+n+q,n_{1}+\bar{n}+q)}^{|n-\bar{n}|}  \nonumber \\
\hspace{-1.8cm} & & \times \chi_{max(n_{2}+n,n_{2}+\bar{n})}^{|n-\bar{n}|} 
 L_{min(n_{1}+n+q,n_{1}+\bar{n}+q)}^{|n-\bar{n}|} (2x^{2}_{1} )
  L_{min(n_{2}+n,n_{2}+\bar{n})}^{|n-\bar{n}|} (2x^{2}_{2}) ,
\end{eqnarray}
with $\chi$ given by equation (\ref{fwignerxi23}). In Fig. 2 we plot the thermal Wigner function of the coherent state of $\mathfrak{su}(1,1)$ for a temperature of $0.005K$.

\newpage

\begin{figure}[!ht]
\centering
\includegraphics[scale=0.08]{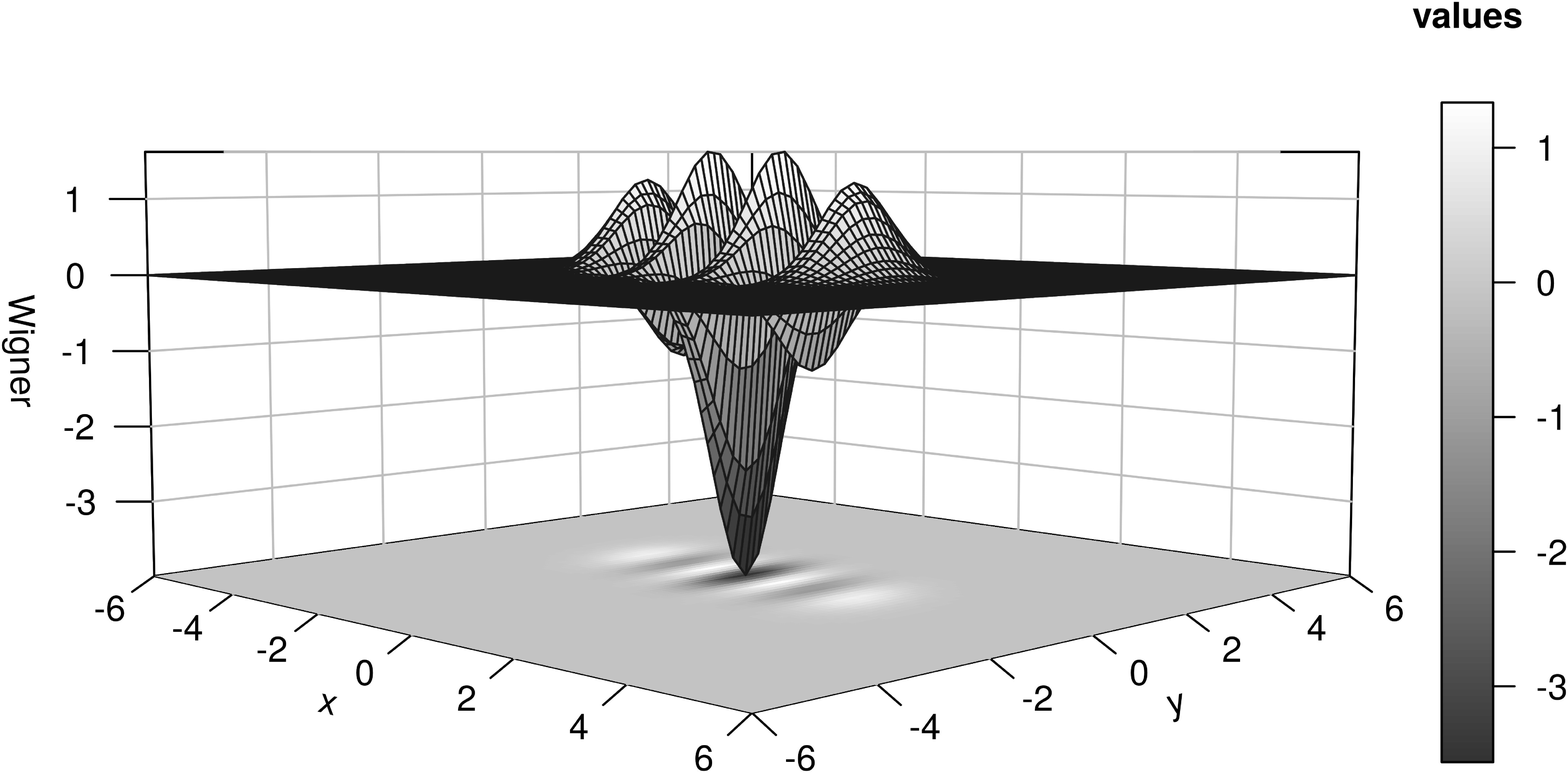} \label{figura-2}
\caption{  Thermal Wigner function for $\mathfrak{su}(1,1)$ Lie algebra with $q=3$, $\zeta=0.1$, $\omega=10^{7}Hz$ and $T=0.005K$.} 
\end{figure}

\section{Conclusions}\label{sec6}

In this paper, we developed and presented the Generalized Thermal Coherent State from coset spaces of Lie groups  perspective, using the Thermofield Dynamics approach. 
This construction allows us to investigate effects of temperature in the Coherent State for an arbitrary Lie algebra for multi-modes. 
As applications we calculated the thermal coherent states associated to $\mathfrak{su}(2)$ and $\mathfrak{su}(1,1)$ Lie algebra and we obtained their thermal density operators. 
Furthermore the Thermal Fidelity and Thermal Wigner Function where obtained. 
The thermal coherent states we obtained reduce to the original pure state in the limit $ T \rightarrow 0 \ \left( \beta \rightarrow + \infty \right) $ for systems with the same phase. 
 Notice that in the framework of the quantum field theory, with continuous limit relation
$\sum_{k}   \rightarrow   \frac{V}{(2\pi)^{3}} \int d^{3}k,$
we have
$\langle \psi(\beta) \vert \psi (\beta') \rangle  \rightarrow  0$ for $ \beta \neq \beta'$,  $V\rightarrow 0$
as thoroughly discussed in the analogous context in the reference \cite{celeghini3}. In the infinity volume limit, there is not unitary operator $U(\beta)$ which maps the Hilbert space onto it self, i. e., the representations are unitarily inequivalent.
As perspectives, an investigation about phase transitions is in progress.

\end{document}